\documentclass[letter]{emulateapj}
\usepackage{graphicx}

\def\j1748{SAX\,J1748.9$-$2021}

\begin{document}

\title{The mass  and the radius of  the neutron star  in the transient
  low mass X-ray binary SAX~J1748.9$-$2021}

\author{Tolga G\"uver\altaffilmark{1}, Feryal \"Ozel\altaffilmark{2,3}}

\altaffiltext{1}{Faculty   of   Engineering  and   Natural   Sciences,
  Sabanc\i~University, Orhanl\i, Tuzla, Istanbul 34956, Turkey}

\altaffiltext{2}{University of  Arizona, Department of  Astronomy, 933
  N. Cherry Ave., Tucson, AZ 85721}

\altaffiltext{3}{Radcliffe  Institute  for   Advanced  Study,  Harvard
  University, 8 Garden St., Cambridge, MA 02138}

\begin{abstract}
  We  use time  resolved  spectroscopy of  thermonuclear X-ray  bursts
  observed from SAX~J1748.9$-$2021 to infer the mass and the radius of
  the neutron star in the  binary. Four X-ray bursts observed from the
  source  with RXTE  enable us  to measure  the angular  size  and the
  Eddington  limit  on  the  neutron  star surface.  Combined  with  a
  distance  measurement to  the  globular cluster  NGC~6440, in  which
  SAX~J1748.9$-$2021 resides, we obtain  two solutions for the neutron
  star  radius and  mass, $R  =  8.18 \pm1.62$  km and  $M=1.78\pm0.3$
  M$_{\sun}$  or   $R=10.93  \pm  2.09$  km  and   $M=  1.33  \pm0.33$
  M$_{\sun}$.
\end{abstract}

\keywords{stars: neutron - X-ray: individual: SAX~J1748.9$-$2021}

\section{Introduction}

The transient neutron-star X-ray binary \j1748 located in the globular
cluster NGC~6440  was discovered with  BeppoSAX during an  outburst in
1998 (in't Zand  et al.\ 1999). The identification of  the optical and
quiescent counterparts of the binary followed shortly afterwards (in't
Zand  et  al.  2001a).   In  2001  and  2005, further  outbursts  were
observed (in't  Zand et  al. 2001b; Markwardt  \& Swank  2005), during
which a number  of X-ray bursts were detected from  \j1748. A total of
16  X-ray  bursts that  were  observed  with  the Rossi  X-ray  Timing
Explorer (RXTE) showed characteristics of thermonuclear flashes on the
neutron star  surface (Galloway et al.\  2008). In this paper,  we use
the time-resolved  spectroscopy of a subsample  of these thermonuclear
bursts to determine the mass and radius of the neutron star in \j1748.

Thermonuclear X-ray bursts are  caused by the unstable nuclear burning
of  matter accreted  onto the  surface of  the neutron  star  from its
companion star.   Helium flashes that  spread across the  neutron star
surface  give  rise to  the  typical sudden  rise  in  the X-ray  flux
accompanied by a  rise in the temperature. In  the cooling phases, the
flux  and the temperature  decrease while  the apparent  emitting area
remains  nearly constant  (see Lewin  et al.\  1993 for  a  review). A
recent analysis by G\"uver et al.\ (2012a) showed that repeated bursts
from the same source give  rise to highly reproducible emitting areas,
strongly  supporting  the  hypothesis  that the  entire  neutron  star
surface is  emitting during  the cooling phase.  Because of  this, the
cooling tails of  X-ray bursts can be analyzed  to measure the angular
size of the neutron star, which can then be converted into an apparent
radius using the distance to the source.

A subset  of the brightest  bursts show Photospheric  Radius Expansion
(PRE), in which the radiative  flux exceeds the local Eddington limit,
causing  the photosphere  to expand  above the  stellar surface  (see,
e.g., Lewin et  al. 1993). This results in  a characteristic evolution
of the  temperature and the  inferred emitting area during  the burst,
which can be  identified using high time  resolution spectroscopy. The
Eddington limit on the stellar surface can then be determined from the
PRE  events  by  measuring  the  flux during  the  expansion  and  the
contraction of the photosphere (Damen et al. 1990).

Thermonuclear  bursts and  PRE  events are  useful  for measuring  the
masses and  radii of  neutron stars  (see, e.g.,  Lewin et  al.\ 1993;
Damen  et  al.\  1990;  Kuulkers  et  al.\  2003;  \"Ozel  2006).   In
particular, a combination of the apparent angular size obtained during
the cooling tails  of X-ray bursts, the Eddington  limit obtained from
the PRE events,  and a measurement of the source  distance can lead to
unique  and uncorrelated  uncertainties in  the stellar  mass $M$  and
radius  $R$. 

A rich burst database consisting of 1187 X-ray bursts observed from 48
neutron-star X-ray binaries  was made available by more  than 10 years
of   RXTE   observations   (Galloway   et  al.\   2008).    The   high
signal-to-noise  ratio, high-time  resolution data  allowed \"Ozel  et
al.\ (2009), G\"uver et al.\ (2010a, b), and \"Ozel et al.\ (2012a) to
measure the masses and radii of four bursting neutron stars and to put
strong observational constraints on the  equation of state cold matter
at supranuclear  densities for the  first time (\"Ozel,  Baym, G\"uver
2010;  see also  Steiner, Lattimer  \& Brown  2010). In  addition, the
frequent recurrence of X-ray bursts  in a large number of neutron-star
X-ray binaries also allowed a  systematic study of the reproducibility
and the uncertainties in the angular sizes and the Eddington limits of
neutron stars measured  using these data sets (G\"uver  et al.\ 2012a,
b). These studies, which included all  X-ray bursters with two or more
PRE  events,   showed  that   systematic  uncertainties  in   the  two
spectroscopic quantities of interest are  limited to 10\% in the whole
sample.

Following the earlier analyses, we  report here the measurement of the
mass and the radius of the  neutron star in the transient X-ray binary
\j1748.  We use the results of G\"uver et al.\ (2012a,b) on four X-ray
bursts, two  of which show evidence  of PRE, observed from  the source
during the  2001 outburst.  (See the  discussion in that paper  and in
Section 2 for the burst selection criteria).

The globular  cluster NGC~6440, which  hosts \j1748, is a  massive and
old Galactic bulge globular cluster  with an age of $11_{-2}^{+3}$~Gyr
(Origlia et  al.\ 2008). Most of  the properties of NGC~6440,  such as
its metallicity and age, are similar to those of 47~Tucanae, which has
been used  as a reference  for globular cluster  distance measurements
(see, e.g.,  Valenti et al.\  2007). NGC~6440  is known to  host other
X-ray  sources (see,  e.g., Pooley  et al.\  2002), which  includes at
least one other transient low-mass  X-ray binary (Heinke et al.\ 2010)
and five radio millisecond pulsars (Freire et al.\ 2008). The distance
to the cluster is well constrained; earlier optical observations yield
$8.4_{-1.3}^{+1.5}$~kpc (Kuulkers 2003) while the near-IR studies give
$\sim 8.2$~kpc (Valenti et al. 2007).

In Section~2, we present the angular size and the Eddington limit from
time resolved X-ray  spectral analysis.  In Section~3, we  make use of
these  along  with the  distance  to  NGC~6440  to find  two  possible
solutions  for the  radius and  mass of  the neutron  star in  \j1748:
$R=10.93 \pm 2.09$  km and $M= 1.33 \pm 0.33$~M$_{\sun}$  or $R = 8.18
\pm  1.62$~km  and $M=1.78  \pm  0.30$~M$_{\sun}$.   In Section~4,  we
present  the  existing  data  on  the  neutron  star  obtained  during
quiescence and discuss the implications of these measurements.

\section{Thermonuclear bursts from \j1748}

We base our  analysis on the data and methods  presented in G\"uver et
al.\ (2012a, b). Those studies analyzed time resolved spectra obtained
by  RXTE  during thermonuclear  bursts  from  all neutron  star  X-ray
binaries that showed two or more PRE events, including \j1748.

RXTE observed \j1748\ for a total  of 149~ks up to June 2007 (Galloway
et al.\  2008).  In  these observations, a  total of  16 thermonuclear
bursts were  detected, all of  which were  within an eight  day period
during  the 2001  outburst of  the  binary. Because  of the  increased
persistent  flux levels  during  the  outburst, all  but  four of  the
observed bursts occurred when the persistent flux was higher than 10\%
of the  peak burst flux  (determined in accordance with  the parameter
$\Gamma$ in Galloway et al.\ 2008).  When the persistent flux is high,
subtracting  the  background accretion  flux  from  the burst  spectra
introduces systematic uncertainties in  the measured surface areas and
Eddington fluxes.   Therefore, as  with all  our previous  studies, we
exclude from the analysis the X-ray bursts that have $\Gamma > 10\%$.

\begin{figure}
\centering
\includegraphics[scale=0.30]{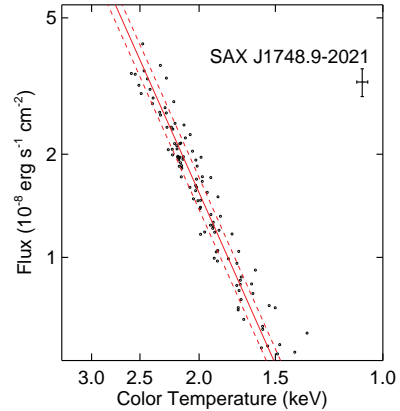}
   \caption{Spectral evolution  during the cooling tails  of the X-ray
     bursts observed  from SAX~J1748.9$-$2021.   Red solid  and dashed
     lines correspond to the most  likely value of the apparent radius
     and to  its systematic uncertainty  and follow $F  \propto T^{4}$
     relation.}
\label{spec_res_1748}
\end{figure}

In G\"uver et al.\ (2012a), we analyzed the spectra during the cooling
tails of these  four bursts, measured the most likely  angular size of
the neutron star, and  quantified any systematic uncertainties arising
from  possible spectral  evolution, inhomogeneous  burning, rotational
broadening, etc.  The  cooling tails of the X-ray bursts  are shown in
Figure~\ref{spec_res_1748}.     Assuming   a    Gaussian   probability
distribution for  the apparent  angular size, we  found that  the most
likely value is  $A = 89.7~({\rm km}/10~{\rm kpc})^2$  with a standard
deviation of $9.6~({\rm km}/10~{\rm kpc})^2$.

\begin{figure}
\centering
\includegraphics[scale=0.30]{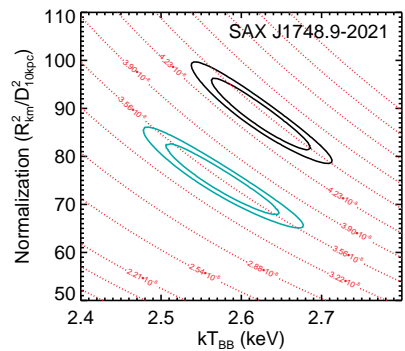}
\caption{68\%   and  90\%   confidence  contours   of  the   blackbody
  normalization and color temperature at the touchdown moment from two
  X-ray bursts that show evidence of photospheric radius expansion.}
\vspace{0.8cm}
\label{td_conf}
\end{figure}

In the burst sample, two events showed evidence of photospheric radius
expansion. We measured the Eddington  flux at the touchdown moment for
these  two  events  (G\"uver  et  al.\ 2012b),  which  we  present  in
Table~\ref{touchdown_fluxes}. We also show in Figure~\ref{td_conf} the
confidence contours of the spectral parameters at the touchdown moment
for each burst.  The fact that  there are only two PRE events prevents
us  from  quantifying  the  level of  systematic  uncertainty  in  the
Eddington flux from these two bursts alone. However, combined with the
sample of  other sources that showed  only a couple of  PRE bursts, we
were  able  to show  statistically  that  the most  likely  systematic
uncertainty in  the Eddington flux  of \j1748\  is equal to  11\% (see
Figures~10 and  11 in Section~6 of  G\"uver et al.\ 2012b).   For this
reason, we adopt a Gaussian probability distribution for the Eddington
flux   of   \j1748\  centered   at   $F_{\rm   Edd}  =   4.03   \times
10^{-8}$~erg~s$^{-1}$~cm$^{-2}$  and  with  a width  of  $0.44  \times
10^{-8}$~erg~s$^{-1}$~cm$^{-2}$.

\begin{deluxetable}{ccc}
  \tablecolumns{3} 
\tablecaption{Touchdown Flux Values.}  
\tablewidth{240pt}
  \tablehead { Burst ID & MJD & Touchdown Flux  \\
  &  & (10$^{-8}$ erg s$^{-1}$ cm$^{-2}$) } 
\startdata
1   & 52190.38947 &  4.52$\pm$0.14 \\
2   & 52190.46882 & 3.54$\pm$0.12 
\enddata
\label{touchdown_fluxes}
\end{deluxetable}

There are sources of possible  systematic uncertainties in addition to
those discussed  in G\"uver et  al.\ (2012a,b) in the  measurements of
the apparent  angular size and  the Eddington flux  from thermonuclear
bursts.  For example,  uncertainties may arise from  anisotropy in the
bursts,  as noted  by  Zamfir  et al.\  (2012)  for  the X-ray  binary
GS~1826$-$24. This is  unlikely to apply to \j1748  given the measured
5\%  systematic variation  in the  radius of  the emitting  region.  A
second uncertainty comes  from the absolute calibration  of the fluxes
measured by  X-ray satellites.  Cross  calibration of RXTE  with other
X-ray  satellites  help in  mitigating  this  uncertainty but  do  not
eliminate   it   (see   Jahoda   et   al.\   2006   and   Steiner   et
al.\ 2010b). Finally, we make here  the assumption that the burst flux
at the  touchdown moment  in PRE bursts  corresponds to  the Eddington
flux on the  neutron star surface. Steiner et  al.\ (2010a) considered
the possibility  that the photosphere  remains above the  neutron star
surface  at  "touchdown" and  discussed  uncertainties  in the  radius
measurements that may arise from relaxing this assumption.

\section{The Mass and Radius of \j1748}

The  first  measurement  of  the  distance  to  the  globular  cluster
NGC~6440, which  was carried out  in the optical wavebands,  yielded a
distance of  $8.4^{+1.5}_{-1.3}$~kpc (Kuulkers et al.\  2003).  A more
recent study  by Valenti et  al.\ (2007)  in the near-IR  obtained the
cluster   distance   by   comparing  the   luminosity   function   and
color-magnitude diagrams  with a  reference cluster, 47  Tucanae. They
found a  distance of $8.2 \pm  0.6$~kpc, which is consistent  with the
earlier measurement  but has  an improved  uncertainty, and  where the
quoted  error   takes  into   account  the   systematic  uncertainties
introduced by the method of comparing to the reference cluster. In the
following, we adopt this latter distance and its uncertainty.

To  determine  the  mass  and  the  radius  of  the  neutron  star  in
SAX~J1748.9$-$2021, we follow the  Bayesian method discussed in detail
in \"Ozel  et al.\ (2009, 2012a).   For each observable, we  assign an
independent probability distribution function.  For the touchdown flux
$P(F_{TD})dF_{TD}$,  the  apparent  angular  size  $P(A)dA$,  and  the
distance $P(D)dD$, we use  Gaussian probability distribution functions
with the parameters discussed above.

There are  two other model parameters  that are needed to  convert the
observables to a measurement of the  neutron star mass and radius: the
hydrogen mass fraction $X$ in  the atmosphere and the color correction
factor $f_{c}$. The  latter is used to convert  the color temperatures
obtained  from  blackbody  fits  to X-ray  spectra  to  the  effective
temperature  of the  neutron star  atmosphere. In  the absence  of any
prior  knowledge, we  assume  for  the hydrogen  mass  fraction $X$  a
box-car distribution  between the entire  range of possible  values of
0.0  and  0.7.   The  color  correction  factor  obtained  from  model
atmospheres  was discussed  in detail  in G\"uver  et al.\  (2010a, b,
2012a).  For this model parameter,  we adopt a box-car distribution in
the range $f_{c}  = 1.35 \pm 0.05$  as in the earlier  studies that is
appropriate for the fluxes observed in the cooling tails of the bursts
from \j1748.

\begin{figure*}
\centering
   \includegraphics[scale=0.40]{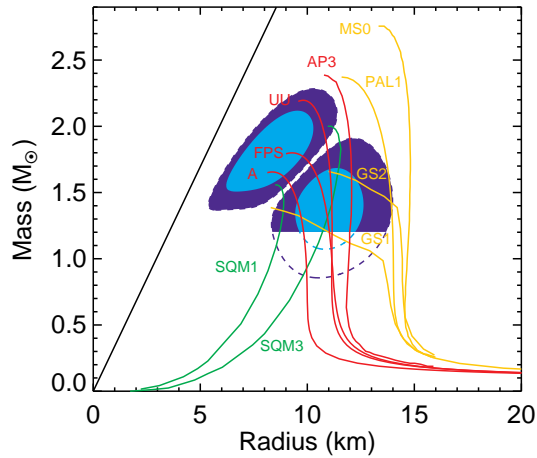} 
   \caption{ Dashed  lines show the 68\% and  95\% confidence contours
     for   the  mass   and  the   radius  of   the  neutron   star  in
     SAX~J1748.9$-$2021. Filled regions  show the more astrophysically
     relevant regions, with M $>$ 1.2 M$_{\sun}$.}
\label{mr_res}
\end{figure*}

We  convert the  likelihoods  over the  touchdown  flux, the  apparent
angular size,  and the  distance to  one over  the neutron  star mass,
radius, and  distance following \"Ozel  et al.\ (2012a)  and integrate
over the  distance likelihood. The resulting  probability distribution
over the  neutron star mass and  radius peaks in two  close regions in
the mass-radius space, as shown in Figure~\ref{mr_res}. Even though we
calculate the full probability  distribution over these parameters, we
show  with filled  contours  the  regions above  $M>  1.2 M_\odot$  in
Figure~\ref{mr_res}, which are astrophysically  more relevant. This is
supported both  observationally, as the vast  majority of measurements
point  to neutron  star masses  above this  limit, and  by studies  of
pre-supernova  stellar  cores,  which  are expected  to  leave  behind
neutron stars with  birth masses $\gtrsim 1.1 M_\odot$  (see \"Ozel et
al.\ 2012b  for a discussion of  all the current measurements  and the
theoretical expectations).

To obtain the individual measurements  of the stellar radius and mass,
we marginalize the two-dimensional  probability distribution over mass
and radius,  respectively.  We find  that the  radius and mass  of the
neutron  star is  either $R  = 8.18  \pm 1.62$~km  and $M  = 1.78  \pm
0.30$~M$_{\sun}$   or  $R=10.93   \pm  2.09$~km   and  $M=   1.33  \pm
0.33$~M$_{\sun}$,  where  the  uncertainties represent  the  1$\sigma$
width of the  best-fit Gaussians to the  probability density functions
over the stellar radius and mass.

\section{Discussion}

Spectroscopic measurements of neutron star radii have been carried out
to date using several techniques.  The first method, which consists of
the  analysis of  thermonuclear bursts  that  we used  in the  present
paper,  has previously  been applied  to four  neutron star  binaries:
EXO~1745$-$248  (\"Ozel  et  al.    2009),  4U~1608$-$52  (G\"uver  et
a.\ 2010a),  4U~1820$-$30 (G\"uver et al.\  2010b), and KS\,1731$-$260
(\"Ozel et  al.  2012a).  Using  the evolution  of burst spectra  as a
function of  burst flux, Zamfir  et al.\  (2012) carried out  a radius
measurement  in  GS\,1826$-$24.  A  second  method  that utilizes  the
thermal X-ray  emission from  neutron star binaries  during quiescence
has  yielded  radius measurements  for  four  additional sources:  the
neutron stars in M13, NGC~2808,  and $\omega$Cen (Webb \& Barret 2007)
as well as U24 in NGC~6397  (Guillot et al.\ 2011; see the compilation
in \"Ozel  (2013) for  a detailed  discussion of  each source  and the
corresponding  measurement   uncertainties  as   well  as   for  radii
measurements in isolated  neutron stars). The mass and  radius we find
here for the  neutron star in SAX~J1748.9$-$2021 is  in good agreement
with these  previous measurements.  Even though  no single measurement
currently constrains at a significant  level the neutron star equation
of state, collectively,  they point to neutron star radii  that are in
the  8-11~km range,  which is  consistent only  with a  subset of  the
proposed equations of state (see \"Ozel, Baym, \& G\"uver 2010).

The two techniques mentioned above have never been applied to the same
source,  which  would  provide  an important  test  for  the  methods'
consistency. For that purpose, X-ray  bursters that are also transient
sources, such as  SAX~J1748.9$-$2021, can be very  useful.  \j1748 has
been in quiescence since its last  outburst in 2005.  During this time
interval, the Chandra X-ray  Observatory observed the globular cluster
NGC~6440 one time in 2009 with an exposure of approximately 50~ks.  We
analyzed  the  data obtained  during  the  observation using  standard
Chandra                          data                         analysis
techniques.\footnote{http://cxc.cfa.harvard.edu/ciao/}   Although  the
source is clearly  detected, with only about 260 source  counts, it is
not  possible  to obtain  any  reliable  constraints on  the  apparent
angular  size  of  the  neutron  star.   However,  unlike  some  other
transients (e.g., Cen~X$-$4, Cackett et al.\ 2010), the existing X-ray
spectrum shows no sign of  a non-thermal component that would indicate
some residual  accretion.  In  that optimal case,  a new  deep Chandra
observation  will provide  the  opportunity to  obtain an  independent
constraint on  the radius  of this  neutron star,  and hence  a direct
comparison of the results of the two methods.

\acknowledgements{We thank  Elena Valenti for very  useful discussions
  on the  measurements of  the distances  to globular  clusters.  This
  work was supported by NASA ADAP  grant NNX10AE89G and in part by the
  Radcliffe Institute  for Advanced Study at  Harvard University. This
  research  has  made  use  of  data obtained  from  the  High  Energy
  Astrophysics Science Archive Research  Center (HEASARC), provided by
  NASA's Goddard Space Flight Center. }

\end{document}